\documentclass[12pt]{article}

\usepackage{multirow}
\usepackage{comment}
\usepackage{epsfig}
\usepackage{epstopdf}
\usepackage{amssymb,amsmath}
\usepackage{comment}

\usepackage{setspace}
\newcommand{\bea}{\begin{eqnarray}}
\newcommand{\eea}{\end{eqnarray}}

\numberwithin{equation}{section}

\begin{document}
\begin{titlepage}
%\begin{flushright}
%
%\end{flushright}
%
\vspace*{10mm}
\begin{center}
\baselineskip 25pt 
{\Large\bf
%%%%%%%%%%%%%%%%%%%%%%%%%%%%%%%%%%%%%%%%%%%%%%%%%%%
Pseudo-Goldstone Dark Matter \\
in gauged $B-L$ extended Standard Model
%%%%%%%%%%%%%%%%%%%%%%%%%%%%%%%%%%%%%%%%%%%%%%%%%%%
}
\end{center}
\vspace{5mm}
\begin{center}
{\large
Nobuchika Okada\footnote{okadan@ua.edu}, 
Digesh Raut\footnote{draut@udel.edu}, 
and Qaisar Shafi\footnote{qshafi@udel.edu}
}
\end{center}
\vspace{2mm}

\begin{center}
{\it
$^{1}$ Department of Physics and Astronomy, \\ 
University of Alabama, Tuscaloosa, Alabama 35487, USA \\
$^{2,3}$ Bartol Research Institute, Department of Physics and Astronomy, \\
 University of Delaware, Newark DE 19716, USA
}
\end{center}
\vspace{0.5cm}
%%%%%%%%%%%%%%%%%%%%%%
\begin{abstract}
%%%%%%%%%%%%%%%%%%%%%%

Gauging the global $B-L$ (Baryon number minus Lepton number) symmetry in the Standard Model (SM)
   is well-motivated since anomaly cancellations require the introduction of three right-handed neutrinos (RHNs)
   which play an essential role in naturally generating tiny SM neutrino masses
   through the seesaw mechanism. 
In the context of the $B-L$ extended SM, we propose a pseudo-Goldstone boson dark matter (DM) scenario 
   in which the imaginary component of a complex $B-L$ Higgs field serves as the DM in the universe. 
The DM relic density is determined by the SM Higgs boson mediated process, 
   but its elastic scattering with nucleons through the exchange of Higgs bosons is highly suppressed 
   due to its pseudo-Goldstone boson nature. 
The model is therefore free from the constraints arising from direct DM detection experiments.    
We identify regions of the model parameter space for reproducing the observed DM density  
   compatible with the constraints from the Large Hadron Collider and the indirect DM searches by Fermi-LAT and MAGIC.
 
%%%%%%%%%%%%%%%%%%%%%%%
\end{abstract}
\end{titlepage}

%%%%%%%%%%%%%%%%%%%%%%%%%%%%%%%%%
\section{Introduction}
\label{sec:Intro}
%%%%%%%%%%%%%%%%%%%%%%%%%%%%%%%%%

According to the widely accepted $\Lambda_{CDM}$ model \cite{Aghanim:2018eyx} around 25\%  
  of the universe's total energy density resides in one or more dark matter (DM) particle.
A neutral weakly interacting massive particle (WIMP), incorporated in new physics 
  beyond the Standard Model (SM), remains an attractive DM candidate. 
The so-called Higgs-portal scalar DM \cite{McDonald:1993ex} is a well-studied WIMP DM scenario, 
  in which a SM singlet real scalar field plays the role of WIMP DM 
  through its renormalizable interaction with the SM Higgs boson. 
Because of its simplicity, the physics of the Higgs-portal scalar DM scenario 
  is determined by only two parameters,  
  a quartic coupling between the scalar DM and the SM Higgs doublet ($\lambda_{HSS}$)
  and the DM mass ($m_S$). 
The constraint from the observed DM relic density determines 
  $\lambda_{HSS}$ as a function of $m_S$, in which case the latter is the unique free parameter of the scenario.

A number of DM detection experiments have been searching for a signal from a DM particle scattering off nuclei. 
No evidence for this has so far been observed, 
  and the most stringent upper bound is reported by  XENON1T experiment \cite{Aprile:2018dbl} and 
  the DarkSide-50 experiment \cite{Agnes:2018ves} for a DM mass $m_S[{\rm GeV}] >  6$ 
  and $1 \leq m_S[{\rm GeV}] \leq6$, respectively. 
For the Higgs-portal scalar DM, the upper bound on DM-nucleon scattering cross section leads to a lower bound 
   on $\lambda_{HSS}$. 
For $1 \, {\rm GeV} < m_S \lesssim$ a few TeV, 
   almost the entire region which can reproduce the observed DM density with $\lambda_{HSS}$ in perturbative regime 
   is excluded, except for a very narrow region in the vicinity of the Higgs boson resonance point 
   of $m_S \simeq m_h/2 \simeq 62.5$ GeV. 
Studies of the Higgs boson decay at the Large Hadron Collider (LHC) \cite{Sirunyan:2018owy}
   exclude the region $m_S < 1$ GeV, which predicts a large invisible branching ratio into a pair of DM particles. 
Although the Higgs-portal scalar DM scenario is relatively straightforward scenario, only a very limited parameter region is allowed.  
See Ref.~\cite{Arcadi:2019lka} for a review on the current status of the Higgs-portal DM scenario.

Recently, a so-called pseudo-Goldstone DM (pGDM) model has been proposed in Ref.~\cite{Gross:2017dan}, 
    which is an extension of the Higgs-portal scalar DM scenario 
    with a (broken) global U(1) symmetry. 
The basic idea is the following: 
It contains a single complex scalar $S$ and its mass term takes the form 
\bea
  \mu_S^2 \left( S^2 + (S^\dagger)^2 \right),  
 \label{ex_mass}  
\eea
  where $\mu_S$ is a real mass parameter. 
In the absence of this term, the model possesses a global U(1) symmetry, which is broken by a nonzero 
  vacuum expectation value (VEV) of the real part of $S$. 
The imaginary component of $S$ (we call it $\chi$) is a massless Nambu-Goldstone (NG) particle 
  in the limit $\mu_S \to 0$. 
Even for $\mu_S \neq 0$, the model has a $Z_2$ symmetry under 
   which $\chi$ has an odd-parity and all the other fields including the SM fields are even. 
Hence, $\chi$ is stable and a Higgs-portal scalar DM candidate.  
A characteristic feature of this model is that despite $\mu_S \neq 0$, 
  $\chi$ retains a Goldstone boson nature with a derivative coupling to the Higgs boson. 
As a result, this coupling disappears in the non-relativistic limit, so that the scattering cross section of the DM particle $\chi$ with a nucleon 
   mediated by the Higgs bosons vanishes \cite{Gross:2017dan}. 
This model is therefore free from the constraints from the direct DM detection experiments.

In this paper we propose a pGDM model 
  based on a simple extension of the minimal $B-L$ (Baryon number minus Lepton number) model \cite{mBL}, 
   where the anomaly-free global $B-L$ symmetry of the SM is gauged. 
We introduce an additional scalar field relative to the minimal $B-L$ model which has a unit $B-L$ charge and whose imaginary component pays the role of pGDM.  
Except for the DM physics, the phenomenology of the model is much the same as that of the minimal $B-L$ model. 
The gauge and mixed gauge-gravitational anomalies are all canceled by the presence of 
   three right-handed neutrinos (RHNs), which acquire their Majorana masses associated 
   with $B-L$ symmetry breaking.  
With the Majorana RHNs and electroweak symmetry breaking, the seesaw mechanism
   works to generate the tiny neutrino masses. 
The model can also account for the observed baryon asymmetry of the universe
   through leptogenesis \cite{LG}.

Our gauge extension of the pGDM model has another theoretical advantage. 
In the original pGDM model \cite{Gross:2017dan}, in order to realize a phenomenologically viable scenario it is essential to introduce the mass squared terms in Eq.~(\ref{ex_mass}) 
   which explicitly break the global U(1) symmetry. 
Since the latter symmetry is not manifest one could, in general, include additional terms. 
However, with such general terms, the DM particle looses its Goldstone boson nature 
   and the model will be severely constrained by the direct DM detection experiments. 
As we discuss in the next section, we effectively realize the terms in  Eq.~(\ref{ex_mass}) 
    after $B-L$ symmetry breaking, and any unwanted terms is forbidden by 
    the $B-L$ symmetry. 
Therefore, we may consider our model as an ultraviolet completion of the original pGDM model.

Unlike the original model in Ref.~\cite{Gross:2017dan} 
   where a $Z_2$ symmetry ensures the stability of the DM particle, 
   the $B-L$ gauge interaction explicitly violates this parity, and hence 
   the DM particle is not entirely stable. 
This fact implies a lower bound on the $B-L$ symmetry breaking scale 
   in order to yield a sufficiently long-lived DM particle. 
Although the pGDM evades the direct detection constraints, 
   we examine constraints on the model parameter space 
   from the LHC and from indirect DM search experiments, 
   such as the Fermi Large Area Telescope (Fermi-LAT) \cite{FermiLAT} and 
   Major Atmospheric Gamma Imaging Cherenkov Telescopes (MAGIC) \cite{magic}. See also Ref.~\cite{fermi+magic}.

This paper is organized as follows: 
In Sec.~\ref{sec:model} we present our pGDM model in the $B-L$ framework. 
We first describe the basic structure of the model, and then show that the DM-nucleon 
   scattering amplitude vanishes in the non-relativistic limit. 
We also estimate the lifetime of the pGDM and obtain a lower bound on the $B-L$ symmetry breaking scale. 
In Sec.~\ref{sec:relic} we identify the parameter region compatible with the observed DM relic density. 
In Sec.~\ref{sec:IDC} we constrain the parameter space of our model by taking into account LHC and indirect DM search experiments. 
Our conclusions are summarized in Sec.~\ref{sec:conc}.

%%%%%%%%%%%%%%%%%%%%%%%%%%%%%%%%%%%%%%
\section{pGDM in $B-L$ extended Standard Model}
\label{sec:model}

%%%%%%%%%%%%%%%%%%%%%%%%%%%%%%%%%%%%%%
\begin{table}[t]
\begin{center}
\begin{tabular}{|c|ccc|c|}
\hline
      &  SU(3)$_c$  & SU(2)$_L$ & U(1)$_Y$ & U(1)$_{B-L}$  \\ 
\hline
$q^{i}_{L}$ & {\bf 3 }    &  {\bf 2}         & $ 1/6$       & $ 1/3 $   \\
$u^{i}_{R}$ & {\bf 3 }    &  {\bf 1}         & $ 2/3$       & $ 1/3 $   \\
$d^{i}_{R}$ & {\bf 3 }    &  {\bf 1}         & $-1/3$       & $ 1/3 $  \\
\hline
$\ell^{i}_{L}$ & {\bf 1 }    &  {\bf 2}         & $-1/2$       & $-1$    \\
$e^{i}_{R}$    & {\bf 1 }    &  {\bf 1}         & $-1$                   & $-1$   \\
\hline
$H$            & {\bf 1 }    &  {\bf 2}         & $- 1/2$       & $0$   \\  
\hline
$N^{i}_{R}$    & {\bf 1 }    &  {\bf 1}         &$0$                    & $-1 $     \\
$\Phi_A$            & {\bf 1 }       &  {\bf 1}       &$ 0$                  & $+2$  \\ 
$\Phi_B$            & {\bf 1 }       &  {\bf 1}       &$ 0$                  & $-1$  \\ 
\hline
\end{tabular}
\end{center}
\caption{
The particle content of our $B-L$ extended SM. 
In addition to the SM particle content ($i=1,2,3$), we have three RHNs ($N_R^i$) 
  and two $B-L$ Higgs fields ($\Phi_{A,B}$). 
The model reduces to the minimal $B-L$ model if we omit $\Phi_B$.
}
\label{tab:1}
\end{table}
%%%%%%%%%%%%%%%%%%%%%%%%%%%%%%%%%%%%%%%%%%%%%%%

We consider a $B-L$ extension of the SM that incorporates a pGDM particle. 
The field content is listed in Table \ref{tab:1}.\footnote{
A $B-L$ model with the same particle content has been investigated before. 
In Ref.~\cite{Okada:2018xdh}, the first order phase transition of the $B-L$ gauge symmetry breaking 
   which generates stochastic gravitational waves has been investigated. 
In Ref.~\cite{Okada:2019sbb}, a scalar DM scenario with vanishing $\Phi_B$ VEV has been discussed. 
} 
In addition to the SM fields, 
  the model includes three right-handed neutrinos ($N_R^i$) in order to cancel all the gauge and mixed gauge-gravitational anomalies.  
The scalar sector includes two new SM singlet Higgs fields, $\Phi_{A}$ and $\Phi_{B}$, 
  with $B-L$ charges $+2$ and $-1$, respectively. 
This charge assignment for $\Phi_{A,B}$ is crucial for incorporating a pGDM particle. 
Note that the model reduces to the minimal $B-L$ model 
  if we omit the new Higgs field $\Phi_B$.

The SM Yukawa sector for the RHNs is extended to include in the Lagrangian density the following terms,  
\bea  
   {\cal L} \supset   -\frac{1}{2} \sum_{i,j=1}^{3} Y_D^{ij} {\overline \ell^i} H N_R^j  
   - \frac{1}{2} \sum_{i=1}^{3} Y_N^i  \Phi_A  \overline{N_{R}^{i~C}} N^i_{R}  +{\rm h.c.},
\eea 
where we have assumed a diagonal basis for the Majorana Yukawa couplings. 
After the electroweak and $B-L$ symmetry breaking, the Dirac and the Majorana masses for the RHNs are generated, 
\bea 
m_D^{ij}= \frac{Y_D^{ij}}{\sqrt{2}} v_{H}, \; \; 
m_{N^i}= \frac{1}{\sqrt{2}} Y_N^i  v_{A},
\label{masses}
\eea
where $v_{H} = \sqrt{2} \langle H^0 \rangle =246$ GeV is a VEV of the charge neutral component ($H^0$) of the SM Higgs doublet, 
  and $v_{A}= \sqrt{2} \langle \Phi_A \rangle$. 
%Also associated with the gauge symmetry breaking, two new scalar fields in addition to the SM Higgs boson, 
%a pseudo-Goldstone particle, and a new $B-L$ gauge boson ($Z^\prime$) acquire their masses 
%which we discuss in the following sub-section. 

%%%%%%%%%%%%%%%%%%%%%%%%%%%%%%
\subsection{Realizing pGDM}
%%%%%%%%%%%%%%%%%%%%%%%%%%%%%%
Let us consider the scalar sector of the model. 
The gauge invariant and renormalizable scalar potential for $\Phi_{A,B}$ and $H$ is given by 
\bea
V = &-&\mu_H^2 \left(H^\dagger H\right) -\mu_A^2 \left(\Phi_A^\dagger \Phi_A\right) -\mu_B^2 \left(\Phi_B^\dagger \Phi_B\right) 
\nonumber \\
&+&  \lambda_{H} \left(H^\dagger H\right)^2 + \lambda_{HA} \left(H^\dagger H\right)\left(\Phi_A^\dagger \Phi_A\right) +\lambda_{HB}\left(H^\dagger H\right)\left(\Phi_B^\dagger \Phi_B\right)  
\nonumber \\
&+& \lambda_{AB} \left(\Phi_A^\dagger \Phi_A\right)\left(\Phi_B^\dagger \Phi_B \right)
+ \lambda_{A} \left(\Phi_A^\dagger \Phi_A\right)^2 
+ \lambda_{B} \left(\Phi_B^\dagger \Phi_B\right)^2  
\nonumber\\
&-& \sqrt{2} \Lambda \left(\Phi_A \, \Phi_B^2 + {\text{h.c}} \right),  
\label{eq:potABH} 
\eea
where $\mu_{A,B,H}$ , $\Lambda$, and quartic scalar coupling parameters ($\lambda_i$) are all real parameters 
  with mass dimension $2, 1$, and $0$, respectively.\footnote{
Although $\Lambda$ can, in general, be complex, 
   it can always be made real by a phase rotation of $\Phi_A$.
} 
This scalar potential is invariant under transformation $\Phi_{A,B} \to  \Phi_{A,B}^\dagger$. 
This indicates that the real components of $\Phi_{A,B} $ are ${Z}_2$-even 
  (${\rm Re}[\Phi_{A,B}] \to {\rm Re} [\Phi_{A,B}]$), 
  while their imaginary components are ${Z}_2$-odd
   (${\rm Im}[\Phi_{A,B}] \to - {\rm Im}[\Phi_{A,B}]$).\footnote{
Instead of $\Phi_{A,B} \to  \Phi_{A,B}^\dagger$, one can consider 
   $\Phi_A \to +\Phi_A^\dagger$ and $\Phi_B \to - \Phi_B^\dagger$.  
However, there is no essential difference in physics. 
We can exchange the role of ${\rm Re}[\Phi_B]$ and ${\rm Im}[\Phi_B]$. 
}
Arranging suitably the parameters in the scalar potential, 
   we obtain the $B-L$ symmetry breaking by $\langle {\rm Re}[\Phi_{A,B}] \rangle \neq 0$. 
After this breaking, a linear combination of ${\rm Im}[\Phi_A]$ and ${\rm Im}[\Phi_B]$    
   forms the would-be Nambu-Goldstone (NG) mode which is eaten by the the $B-L$ gauge boson ($Z^\prime$).  
Its orthogonal combination is a physical massive scalar which, as we will see below, is the desired pGDM particle. 
Note that the covariant derivatives for $\Phi_{A, B}$ explicitly break the symmetry $\Phi_{A,B} \to  \Phi_{A,B}^\dagger$
   and so the pGDM is not stable. 
We will discuss its lifetime later.

Let us first consider the mass spectrum of the model. 
We express the scalar fields as  
\bea
\Phi_{A,B} &=& \frac{1}{\sqrt{2}}\left(\phi_{A,B} + v_{A,B} + i \chi_{A,B}\right)
\nonumber \\
H^0 &=& \frac{1}{\sqrt{2}}\left(h + v_{H}\right), 
\label{eq:phiABH} 
\eea
where $v_{B}= \sqrt{2} \langle \Phi_B \rangle$.   
The stationary conditions around the VEVs lead to 
\bea
\mu_A^2 &=& \lambda_A v_A^2 - \frac{\Lambda v_B^2}{v_A}  + \frac{1}{2} \lambda_{AB} v_B^2+ \frac{1}{2} \lambda_{HA} v_H^2, 
\nonumber \\
\mu_B^2 &=& \frac{1}{2}\left(\lambda_{AB}v_A^2  - 4 \Lambda v_A + 2 \lambda_B v_B^2 + \lambda_{HB} v_H^2\right), 
\nonumber \\
\mu_H^2 &=& \frac{1}{2}\left(2\lambda_H v_H^2 + \lambda_{HA}v_A^2 + \lambda_{HB}v_B^2\right). 
\label{eq:musqABH}
\eea
Substituting Eqs.~(\ref{eq:phiABH}) and (\ref{eq:musqABH}) into Eq.~(\ref{eq:potABH}), 
   we obtain the mass matrices for the real and imaginary components, respectively.  
Since the $Z_2$ symmetry is manifest for the scalar potential, 
   there is no mixing between the real and imaginary components. 
For the real components, the mass matrix is given by 
\bea
V  \supset 
\frac{1}{2}
\begin{bmatrix}
\phi_A & \phi_B & h
\end{bmatrix}
\begin{bmatrix} 
 \Lambda \frac{v_B^2}{v_A}  + 2 \lambda_A v_A^2 &  v_B\left(-2\Lambda + v_A \lambda_{AB}\right) & \lambda_{HA} v_A v_H 
\\
v_B\left(-2\Lambda + v_A \lambda_{AB}\right)  & 2 \lambda_B v_B^2 & \lambda_{HB} v_B v_H 
\\
 \lambda_{HA} v_A v_H  &   \lambda_{HB} v_B v_H  & 2 \lambda_H v_H^2
\end{bmatrix} 
\begin{bmatrix} 
\phi_A \\ \phi_B \\ h
\end{bmatrix}, 
\label{eq:massmatrixABh}
\eea
and the corresponding imaginary component mass matrix is given by  
\bea
V  \supset 
\frac{1}{2}
\begin{bmatrix}
\chi_A & \chi_B
\end{bmatrix}
\begin{bmatrix} 
 \Lambda \frac{v_B^2}{v_A} &  2\Lambda v_B \\ 
 2 \Lambda v_B & 4 \Lambda v_A
\end{bmatrix} 
\begin{bmatrix} 
\chi_A \\ \chi_B 
\end{bmatrix}. 
\label{eq:massmatrixAB}
\eea
We first diagonalize the mass matrix for the imaginary components, 
\bea
\begin{bmatrix} 
\chi_A \\ \chi_B
\end{bmatrix}
=
\begin{bmatrix} 
-\cos\theta &   \sin\theta \\ 
\sin\theta & \cos\theta  
\end{bmatrix} 
\begin{bmatrix} 
{\chi_1} \\ {\chi_2}
\end{bmatrix},
\label{eq:eigenstate} 
\eea
where $\chi_{1,2}$ are the mass eigenstates, and %the mixing angle $\theta$ is expressed as  
%\bea
%\tan2\theta = \frac{2 \left(\frac{v_B}{2 v_A}\right)}{1-\left(\frac{v_B}{2 v_A}\right)^2}. 
%\label{eq:tan2th}
%\eea
\bea
\sin\theta = \frac{v_B}{\sqrt{4v_A^2+v_B^2}}\; , 
\qquad
\cos\theta = \frac{2 v_A}{\sqrt{4v_A^2+v_B^2 }}. 
\label{eq:theta}
\eea
The mass eigenvalues of $\chi_{1,2}$ are given by 
\bea
m_1^2 &=& 0 ,
\nonumber \\
m_2^2 &=& 4\Lambda v_A \left(1+\tan^2\theta\right). 
\label{eq:NGmass}
\eea
%Using the eigenvalues to evaluate $\chi_{1,2}$ and comparing it with the ansatz in Eq.~(\ref{eq:eigenstate}), we obtain 
%From the normalization the eigenvectors $\chi_{1,2}$ we obtain 
%\bea
%\sin\theta = \frac{v_B}{\sqrt{4v_A^2+v_B^2}}\; , 
%\qquad
%\cos\theta = \frac{2 v_A}{\sqrt{4v_A^2+v_B^2 }}. 
%\label{eq:theta}
%\eea
%This is consistent with the $\tan\theta = v_B/(2v_A)$ obtained from Eq.~(\ref{eq:tan2th}). 

In the following, we employ the $R_\xi$-gauge to show that $\chi_1$ is the would-be NG mode absorbed  
 by the gauge boson $Z^\prime$, and $\chi_2$ is the pGDM.  
The kinetic terms for the scalars and the gauge field are given by  
\bea
{\mathcal L} \supset
\left(D^A_\mu \Phi_A\right)^\dagger \left(D^{A\mu} \Phi_A\right) 
+ \left(D^B_\mu \Phi_B\right)^\dagger \left(D^{B\mu}\Phi_B\right) -{\cal Z}^\prime_{\mu \nu} {\cal Z}^{\prime \mu \nu}
-\frac{1}{2 \xi}\left(\partial_\mu {Z^\prime}^\mu + \xi \gamma \chi_1 \right)^2. 
\label{eq:LagKin}
\eea
Here, $D^{A,B}_\mu ={\partial}_u - i g Q_{A,B} Z^\prime_\mu$ is the covariant derivative for $\Phi_{A,B}$ 
  with $Q_{A} = +2$ and $Q_{B} =-1$, respectively, 
  ${\cal Z}^\prime_{\mu \nu}$ is the $Z^\prime$ boson field strength, 
  $\xi$ is the gauge fixing parameter, and  
  $\gamma =g (2 v_A \cos\theta +  v_B \sin\theta) = g \sqrt{4v_A^2+v_B^2}$. 
%  , where $\theta$ defined by Eq.~(\ref{eq:theta}). 
The choice of $\gamma$ eliminates the mixing terms $\chi_{1,2} (\partial^\mu Z^\prime)$.  
We rewrite Eq.~(\ref{eq:LagKin}) in terms of the mass eigenstates as 
\bea
{\cal L} &\supset& 
\frac{1}{2} 
{Z^\prime}^\mu \left(
\eta_{\mu\nu}\partial_\alpha \partial^\alpha  - \left(1-\frac{1}{\xi}\right) \partial_\mu \partial_\nu 
\right) {Z^\prime}^\nu 
+  \frac{1}{2} \gamma^2 {Z^\prime}_\mu  {Z^\prime}^\mu  
\nonumber \\
&+& \frac{1}{2} \left(\partial_\mu \chi_1\right) \left(\partial^\mu \chi_1\right) - \frac{1}{2} \xi \gamma^2 \chi_1^2 
+ \frac{1}{2} \left(\partial_\mu \chi_2\right) \left(\partial^\mu \chi_2\right) - \frac{1}{2} m_2^2 \chi_2^2  
\nonumber \\
&-&  2 g \left(-2 \sin\theta \phi_A + \cos\theta \phi_B\right)\left( \partial_\mu \chi_2\right) {Z^\prime}^\mu. 
\label{eq:kinlag}
\eea
Here, as usual in $R_\xi$-gauge, 
  $\gamma$ is identified with the $B-L$ gauge boson mass, $\gamma = m_{Z^\prime}$,  
  and $\chi_1$ is the would-be NG mode whose mass squared is given by $\xi m_{Z^\prime}^2$. 
In the following, we employ the unitary gauge ($\xi \to \infty$), such that 
  the would-be NG mode $\chi_1$ decouples from the system. 
The last line of Eq.~(\ref{eq:kinlag}) shows that 
   the ${Z_2}$ parity is not manifest in the gauge sector, 
   and $\chi_2$ decays through this triple coupling. 
In the next subsection, we estimate the lifetime of $\chi_2$. 
As expected, if $Z^\prime$ and $\phi_{A,B}$ are sufficiently heavy, 
   $\chi_2$ can be sufficiently long-lived in order to be a viable DM in the universe.

%%%%%%%%%%%%%%%%%%%%%%%%%%%%%%%%%%
\subsection{pGDM Direct Detection Amplitude and Lifetime}
%%%%%%%%%%%%%%%%%%%%%%%%%%%%%%%%%%

To check if the elastic scattering cross section of the pGDM ($\chi_2$) 
  with nucleons is adequately suppressed 
  and its lifetime is long enough, 
  let us first consider the so-called ``spurion'' limit. 
In this limit, we take 
  $v_A \gg v_B, v_H$, $\lambda_A v_A^2 \gg \Lambda v_B$, and $\lambda_{AH}, \lambda_{AB} \to 0$, 
   so that the mass matrix of Eq.~(\ref{eq:massmatrixABh}) becomes block-diagonal 
   and $\phi_A$ is decoupled from the system. 
We have $\theta \ll 1$ in Eq.~(\ref{eq:eigenstate}) for $v_A \gg v_B$, 
   and thus $\chi_1 \simeq -\chi_A$ and the pGDM $\chi_2 \simeq \chi_B$. 
Therefore, in the spurion limit (and in the unitary gauge), 
   $\Phi_A$ looses its dynamical degrees of freedom and works as an external field
   with $\langle \Phi_A \rangle$. 
Next, we consider the following mass matrix for $\phi_B$ and $h$: 
\bea
V  \supset 
\frac{1}{2}
\begin{bmatrix}
 \phi_B & h
\end{bmatrix}
\begin{bmatrix} 
 2 \lambda_B v_B^2 & \lambda_{HB} v_B v_H 
\\
 \lambda_{HB} v_B v_H  & 2 \lambda_H v_H^2
\end{bmatrix} 
\begin{bmatrix} 
 \phi_B \\ h
\end{bmatrix}. 
\label{eq:massmatrixHB}
\eea
If we ignore the $B-L$ gauge interaction,  
    the spurion limit effectively realizes the original pGDM model.

The elastic scattering of pGDM ($\chi_B$) with nucleons is  mediated by 
   two Higgs bosons which are linear combinations of $h$ and $\phi_B$. 
The amplitude of the scattering is readily evaluated in the flavor basis.   
%The relevant interaction terms of the pGDM with $h$ and $\phi_B$ are given by 
%\bea
%{\cal L} \supset \left(\lambda_B v_B \phi_B  + \frac{1}{2}\lambda_{HB} v_H h \right) \chi_B \chi_B.
%\label{eq:DMint} 
%\eea
%
The relevant terms for this analysis are given by 
\bea
{\cal L} \supset  S^\dagger ( \Box + M_S) S + C_{SBB} S \chi_B^2 +   C_{hff} S {\bar f}_{SM} f_{SM},  
\eea
where $S \equiv (\phi_B, h)^{\rm T}$, $M_S$ is the $2 \times 2$ mass matrix defined in Eq.~(\ref{eq:massmatrixHB}), 
  $C_{SBB} = (\lambda_B v_B, \lambda_{BH} v_H/2)$, 
   and the last term is the Yukawa interaction of $h$ with SM fermions 
   with $C_{hff} = Y_{h f{\bar f}}(0,1)$. 
Now we can express the scattering amplitude as  
\bea
{\cal M} \propto C_{BBS}\; \frac{1}{t - M_S}\; C_{hff}^{\rm T}. 
\eea
Since this scattering occurs at very low energies, 
   the zero momentum transfer limit of $t\to 0$ is a good approximation: 
\bea
{\cal M} (t \to 0) &\propto& C_{BBh}\; M_S^{-1}\; C_{hff}^{\rm T},
\nonumber \\
&\propto& 
\begin{bmatrix}
\lambda_B v_B & \lambda_{HB} v_H/2
\end{bmatrix}
\begin{bmatrix} 
 2 \lambda_H v_H^2 & -\lambda_{HB} v_B v_H 
\\
- \lambda_{HB} v_B v_H  & 2 \lambda_B v_B^2
\end{bmatrix} 
\begin{bmatrix} 
0 \\ 1
\end{bmatrix} = 0. 
\eea
Therefore, the pGDM scattering amplitude vanishes in the $t\to 0$ limit.

Before moving on to a more general analysis for the pGDM scattering amplitude 
   by taking $\phi_A$ into account, 
let us estimate the pGDM lifetime in the spurion limit.  
The pGDM decays through the interaction, 
\bea
{\cal L} \supset - 2 g \phi_B \left( \partial_\mu \chi_B\right) Z^{\prime \mu}. 
\label{ssz}
\eea
As an example, 
we consider a pGDM mass of $m_{DM} \sim 100$ GeV. 
Since both the $Z^\prime$ boson and $\phi_B$ have couplings with SM fermions (the latter through its mixing with the SM Higgs boson), 
    the main decay mode is $\chi_B \to {Z^\prime}^* \phi_B^* \to {\bar f}_{SM} f_{SM} {\bar f}_{SM} f_{SM}$ 
    through off-shell $\phi_B$ and $Z^\prime$,
   where $f_{SM}$ represents a SM fermion. 
We estimate the pGDM lifetime to be 
\bea
  \tau_{DM} \simeq \frac{(10 \pi)^5}{Y_b^2 \sin^2\theta_H}
   \left(\frac{v_A}{m_{DM}}\right)^4\left(\frac{m_B}{m_{DM}}\right)^4 (m_{DM})^{-1},
\eea
  where $Y_b$ is the bottom Yukawa coupling,  
  $m_B$ is the mass of $\phi_B$, 
  and $\sin \theta_H$ quantifies the mixing of $\phi_B$ with the SM Higgs boson.  
If we require a lower bound $\tau_{DM} \gtrsim 10^{26}$ sec from the cosmic ray observations \cite{Essig:2013goa}, 
   we find 
\bea
v_A \gtrsim 1.22 \times 10^{11} {\rm GeV}
\left(\frac{600 \;{\rm GeV}}{m_B}\right) \left(\frac{m_{DM}}{100\; {\rm GeV}}\right)^{9/4} \left( \frac{\sin \theta_H}{0.2}\right)^{1/2}. 
\eea
The stability of the DM particle requires $v_A$ to be at the intermediate scale or higher. 
In our model, we assume a vanishing kinetic mixing between $Z^\prime$ and the SM $Z$ bosons. 
If the mixing which we parametrize as $\epsilon$ exists, the DM particle has an interaction with $Z$ boson
given by Eq.~(\ref{ssz}) with a replacement $Z^{\prime \mu} \to \epsilon Z^\mu$. 
Considering the decay mode of $\chi_B \to Z^* \phi_B^* \to {\bar f}_{SM} f_{SM} {\bar f}_{SM} f_{SM}$, 
we find an upper bound $\epsilon \lesssim (v_H/v_A)^2 = {\cal O} (10^{-18})$ for $v_A ={\cal O}(10^{11})$ GeV.

Let us now calculate the pGDM scattering amplitude for the more general case 
  by taking $\phi_A$ into account. 
In this case, the pGDM is a linear combination of $\chi_A$ and $\chi_B$ as defined in Eq.~(\ref{eq:eigenstate}), 
  and the pGDM scattering with a nucleon is mediated by three Higgs mass eigenstates 
  which are linear combinations of $h$, $\phi_A$ and $\phi_B$. 
Because of the presence of the extra scalar $\phi_A$, the vanishing scattering amplitude 
  for the limit of $t \to 0$ is not guaranteed.  
We work in the flavor basis with $S = (\phi_A, \phi_B, h)^{\rm T}$, and the relevant terms are given by 
\bea
{\cal L} \supset  S^\dagger ( \Box + M_S) S +  C_{hff} S {\bar f}_{SM} f_{SM} 
+ \left( C_{AAS} S   + C_{BBS} S  + C_{ABS} S  \right) \chi_2^2. 
\label{eq:lagDMABh}
\eea
Here, $M_S$ is the $3 \times 3$ mass matrix in Eq.~(\ref{eq:massmatrixABh}), 
   the second term is the interaction of $h$ with the SM fermions 
    $C_{hff} = Y_{h f \bar{f}} \, (0,0,1)$, and
\bea
C_{AAS} &=& \sin^2\theta \left(2\lambda_A v_A,\; \lambda_{AB} v_B,\;  \lambda_{HA} v_H\right), 
\nonumber \\
C_{BBS} &=&  \cos^2\theta \left(2\Lambda + \lambda_{AB}  v_A, \;  2 \lambda_{B}  v_B, \;   \lambda_{HB} v_H\right), 
\nonumber \\
C_{ABS} &=&\sin\theta \cos\theta \left(0,\;2\Lambda, \; 0\right). 
\eea 
%where we have extracted the coupling of DM pair with $S$ field from its interaction in the flavor basis $\chi_{A,B}$. 
The total amplitude in the limit of $t\to 0$ is expressed as 
\bea
{\cal M} \propto \left(C_{AAS}+ C_{BBS}+ C_{ABS}\right)\; M_S^{-1}\; C_{hff}^{\rm T}. 
\eea

We have previously found that $v_A$ must be higher than the intermediate scale in order 
   to make the pGDM sufficiently long-lived. 
Thus, in order not to significantly alter the SM-like Higgs boson mass eigenvalue from the mass matrix of Eq.~(\ref{eq:massmatrixABh}), 
   we set $\lambda_{AH} \to 0$ in the following analysis. 
The amplitude is then expressed as  
\bea
{\cal M} \propto 
\frac{\Lambda \lambda_{HB} \cos^2\theta 
\left(4 \Lambda + v_A \left(\lambda_A - 2\lambda_{AB}\right) + 2 \tan^2\theta \left( \Lambda + v_A \left(\lambda_A + \lambda_{AB}\tan^2\theta\right)\right)\right)}
{v_H 
\left(
\lambda_H 
\left(\left(\lambda_{AB}^2 v_A -2\Lambda \right)^2 - 4 v_A^2 \lambda_A \lambda_B \right) 
+ v_A^2 \lambda_A \lambda_{HB}^2 
+2 v_A \Lambda \tan^2\theta \left(-4 \lambda_B \lambda_H + \lambda_{HB}\right)
\right)}.
\label{eq:amp1} 
\eea
Because of the perturvativity constraint for the Higgs-portal scalar DM scenario, 
   we are interested in a DM mass ($m_{DM}=m_2$) less than a few TeV. 
 From Eq.~(\ref{eq:NGmass}), we find $\Lambda  \sim m_{DM}^2/v_A \ll 1$.
This simplifies the amplitude formula to  
\bea
{\cal M} \propto
\frac{\Lambda \lambda_{HB} \left(2 \lambda_A - 4 \lambda_{AB}\right)}{2 v_A v_H \left(\lambda_H \lambda_{AB}^2 + \lambda_A \left(\lambda_{HB}^2- 4\lambda _B\lambda_H\right)\right)} %\simeq %\frac{\lambda_{HB} \Lambda }{v_A}, 
  \propto \frac{\Lambda}{v_A}, 
\label{eq:ampfinal}
\eea
which is adequately suppressed. 
To obtain the final expression in Eq.~(\ref{eq:ampfinal}), we have set all the quartic couplings to be of the same order. 
Note that ${\mathcal M} = 0$ cannot be realized even for the momentum transfer $t= 0$.  
This is because the last term in Eq.~(\ref{eq:potABH}), $\Lambda \left(\Phi_A \, \Phi_B^2\right)$, introduces non-derivative coupling between the DM and the scalars and the Goldstone boson nature of the DM particle is lost.

%%%%%%%%%%%%%%%%%%%%%%%%%%%%%%
\section{DM relic density}
%%%%%%%%%%%%%%%%%%%%%%%%%%%%%%
\label{sec:relic}

In this section we numerically evaluate the thermal relic density of the DM particle by solving the Boltzmann equation, 
\bea 
  \frac{dY}{dx}
  = - \frac{\langle \sigma v \rangle}{x^2}\frac{s (m_{2})}{H(m_{2})} \left( Y^2-Y_{EQ}^2 \right). 
\label{eq:Boltzman}
\eea  
Here, $x = m_{2}/T$ is a dimensionless parameter where  $T$ is the temperature of the Universe, 
$H(m_{2})$ is the Hubble parameter and $Y = n/s$ is the DM yield which is defined as the ratio of the DM number density ($n$) to the entropy density ($s$), 
and $Y_{EQ}$ is the yield of the DM particle in thermal equilibrium:  
\bea 
s(m_{2}) = \frac{2  \pi^2}{45} g_\star m_{2}^3,  \; \; 
 H(m_{2}) =  \sqrt{\frac{\pi^2}{90} g_\star} \frac{m_{2}^2}{M_P}, \; \; 
 Y_{EQ}(x) =  \frac{g_{DM}}{2 \pi^2} \frac{x^2 m_{2}^3}{s(m_{2})} K_2(x),   
\eea 
where $K_2$ is the Bessel function of the second kind. 
The thermal average of the total pair annihilation cross section of the DM particles times its relative velocity, $\langle\sigma v\rangle$ in Eq.~(\ref{eq:Boltzman}), can be evaluated as 
\bea 
\langle \sigma v \rangle =  \frac{g_{DM}^2}{64 \pi^4}
  \left(\frac{m_{3}}{x}\right) \frac{1}{n_{EQ}^{2}}
  \int_{4 m_{3}^2}^\infty  ds \; 2 (s- 4 m_{3}^2) \sigma(s) \sqrt{s} K_1 \left(\frac{x \sqrt{s}}{m_{3}}\right),
\label{eq:ThAvgSigma}
\eea
where $g_{DM} (= 1)$ counts the degrees of freedom of the scalar DM particle,  
the equilibrium number density of the DM particle $n_{EQ}=s(m_{2}) Y_{EQ}/x^3$,  $\sigma (s)$ 
is the total DM particle annihilation cross section, and $K_1$ is the modified Bessel function of the first kind.  
The relic density of the DM particle at the present time is evaluated as 
\bea 
  \Omega_{DM} h^2 =\frac{m_{3} s_0 Y(x\to\infty)} {\rho_c/h^2}, 
\eea 
where $s_0 = 2890$ cm$^{-3}$ is the entropy density of the present Universe and $\rho_c/h^2 =1.05 \times 10^{-5}$ GeV/cm$^3$ is the critical density. 
The Planck satellite experiment has measured $\Omega_{DM} h^2 = 0.1200 \pm 0.0012$ \cite{Aghanim:2018eyx}.

In the following, we consider the spurion limit case to be our benchmark, namely, $v_A \gg v_B, v_H$, $\lambda_A v_A^2 \gg \Lambda v_B$, and $\lambda_{AH}, \lambda_{AB} \to 0$. 
In this limit $\phi_A$ is decoupled from the system and the DM mass eigenstate $\chi_2 \simeq \chi_B$. 
The real sector includes $\phi_B$ and $h$ that mix according to the mass matrix in Eq.~(\ref{eq:massmatrixHB}) which we diagonalize by defining the mass eigenstates ${\tilde \phi_B}$ and ${\tilde h}$ as follows: 
\begin{eqnarray}
\begin{bmatrix} 
\phi_B \\h
\end{bmatrix}
=
\begin{bmatrix} 
\cos\theta_H &   -\sin\theta_H \\ 
\sin\theta_H & \cos\theta_H  
\end{bmatrix} 
\begin{bmatrix} 
{\tilde \phi_B}  \\{\tilde h}
\end{bmatrix}. 
\label{eq: eigenstate}
\end{eqnarray} 
Here, the mixing angle $\theta_H$ is determined by  
\bea
2 v_{B} v_{H}  \lambda_{HB}= (m_B^2 -m_h^2) \tan2\theta_H.  
\label{eq: mixings} 
\eea 
The masses of ${\tilde \phi_B}$ and  ${\tilde h}$  are given by
\bea
{m}_{{\tilde B},{\tilde h}}^2 = \frac{1}{2} \left(m_B^2 + m_h^2 \pm \frac {m_B^2 - m_h^2} {\cos2\theta_H}\right), 
\eea
respectively, where $m_h = 2 \lambda_h v_H^2$.
The interaction between the DM and $\tilde h$/${\tilde \phi}_B$ is given by 
\bea
{\cal L} \supset  \frac{{ \lambda _{\tilde h} }\; v_H}{2} \;{\tilde h}\chi_2 \chi_2 + \frac{{ \lambda_{\tilde B}}\; v_H}{2} \;{\tilde \phi_B}\chi_2 \chi_2, 
\label{eq:DMDMS}
\eea
where 
\bea
{ \lambda _{\tilde h} } &=& -\frac{2 {m}_{{\tilde h}}^2}{v_B v_H} \sin\theta_H, \nonumber \\
{ \lambda _{\tilde B} } &=& +\frac{2 {m}_{{\tilde B}}^2}{v_B v_H} \cos\theta_H. 
\eea

To evaluate the DM relic density, let us set $m_B = 600$ GeV and $\sin\theta_H = 0.2$ to be our benchmark. 
In this case, for the DM mass $m_{2} \lesssim200$ GeV, the DM scenario is effectively the same as the Higgs-portal scalar DM scenario such that the DM interaction is given by the first term in Eq.~(\ref{eq:DMDMS}). 
The DM pair annihilation processes and therefore the DM relic abundance is determined by only two free parameters, namely, $m_{2}$ and $\lambda _{\tilde h}$. 
The DM pair annihilation processes include various final states that include SM fermions ($f$), 
   the weak gauge bosons ($W$ and $Z$) and the SM Higgs boson ($h$). 
The DM annihilation cross sections for the various final states are given by \cite{Guo:2010hq}: 
\begin{eqnarray}
{\sigma}_{ff} &=& \sum_f \frac{\lambda _{\tilde h}^2 m_f^2}{\pi}
\frac{1}{(s-m_{{\tilde h}}^2)^2+m_{{\tilde h}}^2 \Gamma_{{\tilde h}}^2}
\frac{(s-4m_f^2)^{\frac{3}{2}}}{\sqrt{s}}, \label{FF} 
\nonumber \\
{\sigma}_{ZZ} &=&  \frac{\lambda _{\tilde h}^2 }{4 \pi}
\frac{s^2}{(s-m_{{\tilde h}}^2)^2+m_{{\tilde h}}^2 \Gamma_{{\tilde h}}^2} \sqrt{1-\frac{4
m_{Z}^2}{s}} \left(1- \frac{4m_{Z}^2}{s}+ \frac{12 m_{Z}^4}{s^2}
\right),
\nonumber \\
{\sigma}_{WW} &=&  \frac{\lambda _{\tilde h}^2 }{2 \pi}
\frac{s^2}{(s-m_{{\tilde h}}^2)^2+m_{{\tilde h}}^2 \Gamma_{{\tilde h}}^2} \sqrt{1-\frac{4
m_{W}^2}{s}} \left(1- \frac{4m_{W}^2}{s}+ \frac{12 m_{W}^4}{s^2}
\right),
\nonumber \\
{\sigma}_{{\tilde h}{\tilde h}} &=&  \frac{\lambda _{\tilde h}^2 }{4 \pi} \sqrt{1-\frac{4
m_{{\tilde h}}^2}{s}} \left[ \left(\frac{s+ 2 m_{{\tilde h}}^2}{s - m_{{\tilde h}}^2}\right)^2
- \frac{16\lambda _{\tilde h} v_H^2}{s-2 m_{{\tilde h}}^2} \frac{s+ 2 m_{{\tilde h}}^2}{s
- m_{{\tilde h}}^2} F(\alpha) 
\right. 
\nonumber \\
&& \left. 
+ \frac{32 \lambda _{\tilde h}^2 v_{H}^4}{(s-2
m_{{\tilde h}}^2)^2} \left( \frac{1}{1-\alpha^2} + F(\alpha)\right) \right],
\label{hh}
%       (2-5)
\label{eq:DMcs}
\end{eqnarray}
where $m_h = 125$ GeV is the SM Higgs boson mass, $s$ is the square of the center-of-mass energy, 
$F(\alpha)\equiv\mbox{arctanh}(\alpha)/\alpha$ with $\alpha
\equiv\sqrt{(s-4m_{{\tilde h}}^2)(s-4m_2^2)}/(s-2m_{{\tilde h}}^2)$, and 
$\Gamma_{{\tilde h}}$ is the total decay width of the SM Higgs boson, 
including ${\tilde h}\to \chi_2 \chi_2$ if allowed by kinematics ($m_{\tilde h} > 2 m_2$),  
\bea
\Gamma_{{\tilde h}} & = & \frac{\sum_f m_f^2 }{8 \pi v_{\rm H}^2}
\frac{(m_{{\tilde h}}^2 - 4 m_f^2)^{3/2}}{m_{{\tilde h}}^2} + \frac{m_{\tilde h}^3}{32 \pi
v_{\rm H}^2} \sqrt{1- \frac{4 m_Z^2}{m_{\tilde h}^2}} \left(1-
\frac{4m_{Z}^2}{m_{\tilde h}^2}+ \frac{12 m_{Z}^4}{m_{\tilde h}^4}\right) \\ \nonumber
& + &  \frac{m_h^3}{16 \pi v_{\rm H}^2} \sqrt{1- \frac{4
m_W^2}{m_{\tilde h}^2}} \left(1- \frac{4m_{W}^2}{m_{\tilde h}^2}+ \frac{12
m_{W}^4}{m_{\tilde h}^4}\right) + \Gamma ({\tilde h}\to \chi_2 \chi_2), 
%       (6)
\label{eq:gammah}
\eea
where,  
\bea
\Gamma({\tilde h}\to \chi_2 \chi_2) =  \frac{\lambda_{\tilde h}^2 v_{\rm H}^2}{32 \pi}
\frac{\sqrt{m_h^2 - 4 m_2^2}}{m_{h}^2}. 
\eea
The total DM annihilation cross section is given by 
\bea
\sigma(s) = \sigma_{ff}+ \sigma_{ZZ} + \sigma_{WW}+ \sigma_{hh}. 
\label{eq:tcs}
\eea

\begin{figure}[t]
\begin{center}
\includegraphics[scale=0.65]{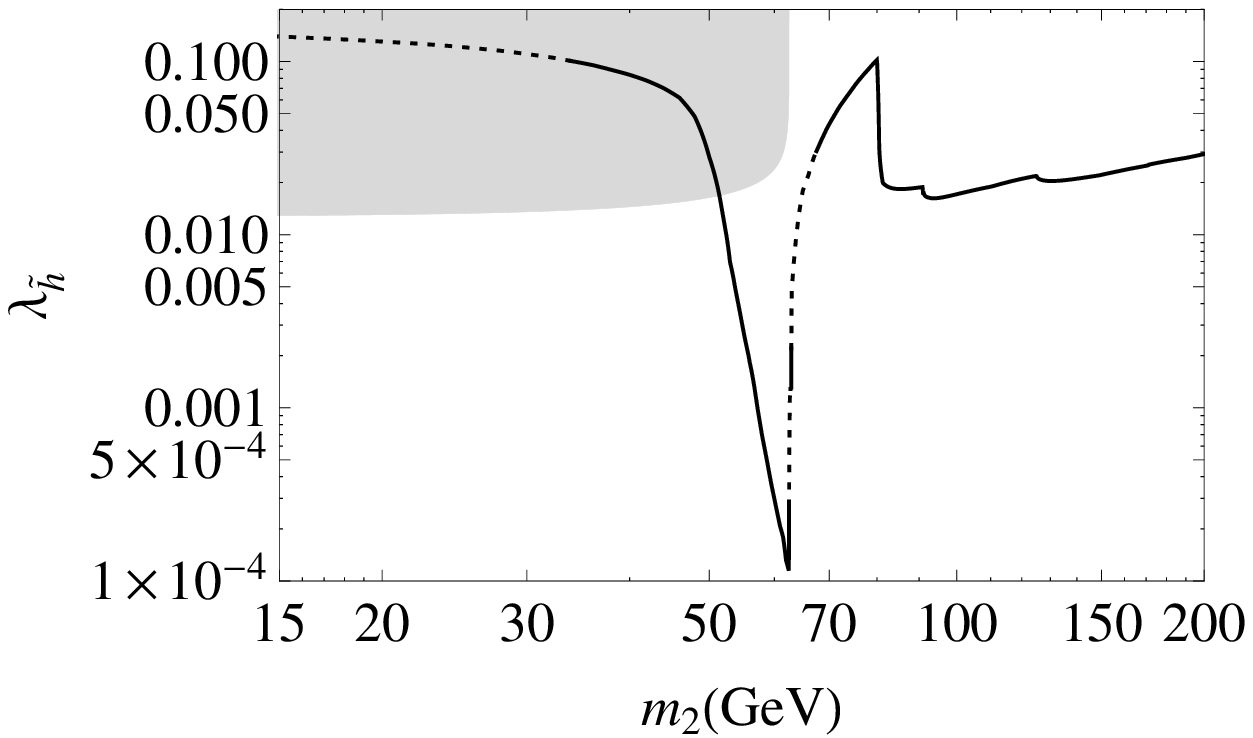}\;
\includegraphics[scale=0.65]{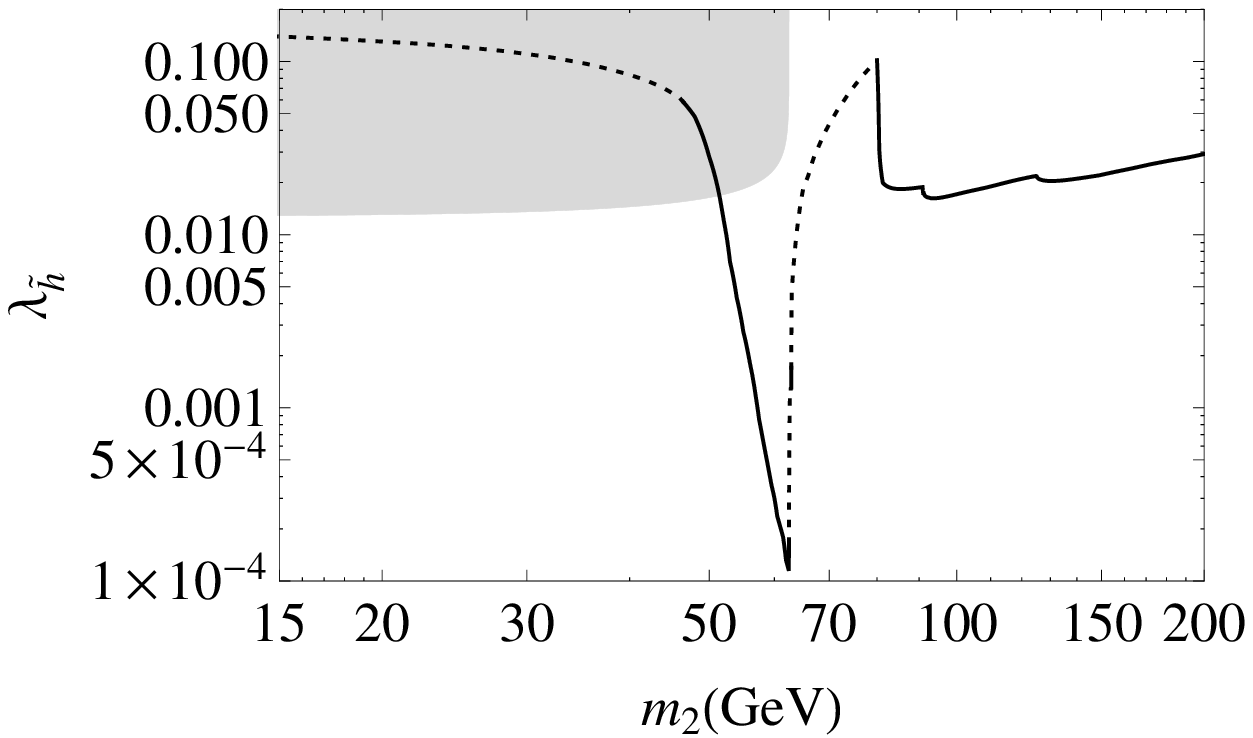}
\end{center}
\caption{
Along the curves the relic abundance constraint is satisfied. 
The dashed region are excluded by Fermi-LAT (left) and Fermi-LAT + MAGIC (right). 
The gray shaded region is excluded by the LHC experiment (see the next section). 
}
\label{fig:relic}
\end{figure}

The DM relic density is controlled by only two free parameters, namely, $m_{2}$ and $\lambda _{\tilde h}$. 
Numerically solving the Boltzmann equation and imposing $\Omega_{DM} h^2 = 0.120$, 
  we have obtained $\lambda _{\tilde h}$ as a function of $m_{2}$ as shown in Fig.~\ref{fig:relic}. 
Here, $\Omega_{DM} h^2 = 0.120$ is reproduced along the curved lines in both panels. 
In the left (right) panel, the dashed region of the curves are excluded by the indirect DM detection constraint from Fermi-LAT (combined Fermi-LAT and MAGIC). 
In both panels, the gray shaded region is excluded by the LHC results 
  on the invisible Higgs boson decay mode, 
$BR({\tilde h} \to \chi_2 \chi_2) \leq 0.16$ \cite{Sirunyan:2018owy}. 
The DM indirect detection and collider search will be discussed in the next section. 

%%%%%%%%%%%%%%%%%%%%%%%%%%%
%%%%%%%%%%%%%%%%%%%%%%%%%%%
%%%%%%%%%%%%%%%%%%%%%%%%%%%
%%%%%%%%%%%%%%%%%%%%%%%%%%%

\section{Indirect Detection and Collider Bounds}
\label{sec:IDC}
%%%%%%%%%%%%%%%%%%%%%%%%%%%
%%%%%%%%%%%%%%%%%%%%%%%%%%%
\begin{figure}[t]
\begin{center}
\includegraphics[scale=0.6]{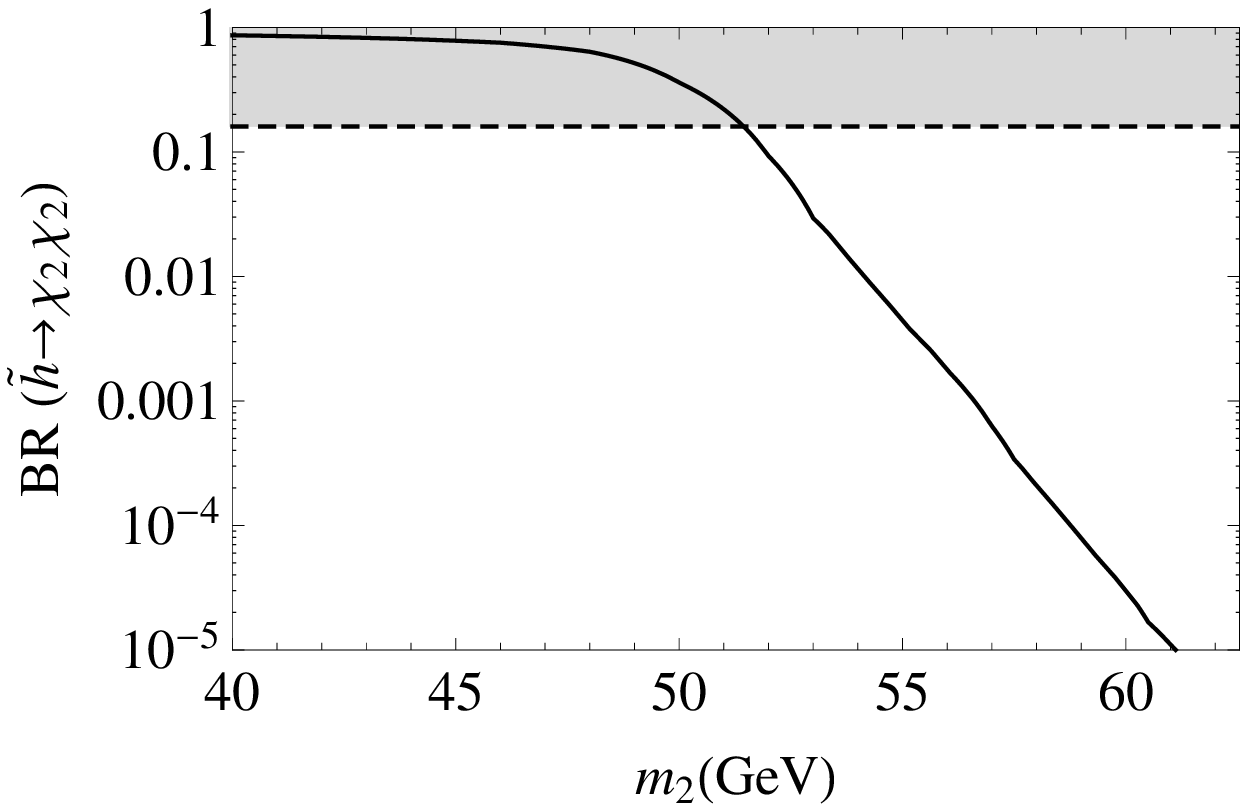}\;
\includegraphics[scale=0.65]{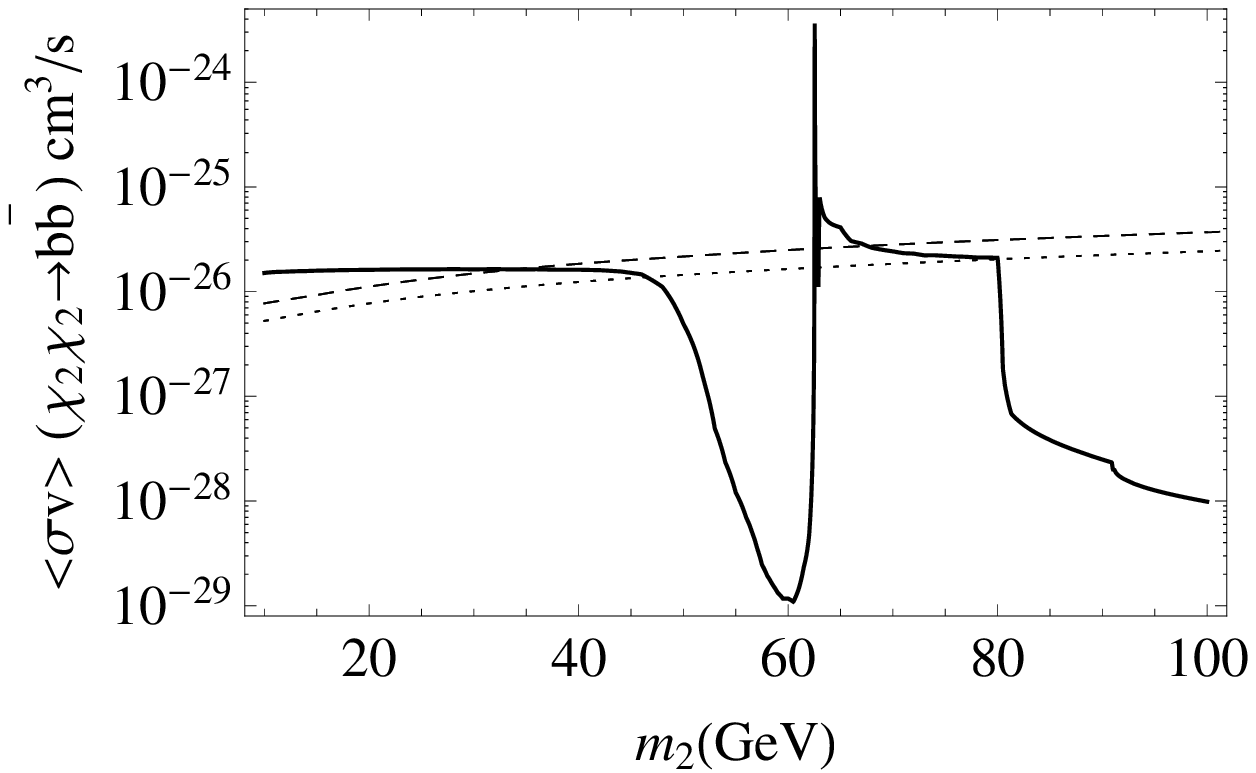}
\end{center}
\caption{
{\it Left panel}: 
Invisible branching ratio for the Higgs boson decay into a pair of pGDMs (solid line) 
   along which $\Omega_{DM} h^2 =0.120$ is reproduced, 
   together with the LHC constraint (gray shaded).  
{\it Right panel}: 
The pGDM pair annihilation cross section into a pair of bottom quarks (solid curve)
  along which $\Omega_{DM} h^2 =0.120$ is reproduced, 
   together with the upper bounds from Fermi-LAT (dashed line) 
   and the combined Fermi-LAT and MAGIC (dotted line).
}
\label{fig:2}
\end{figure}
%%%%%%%%%%%%%%%%%%%%%%%%%%%
%%%%%%%%%%%%%%%%%%%%%%%%%%%

Since the pGDM evades the direct DM detection constraints, we consider the constraints from the LHC and indirect DM detection experiments. 
Let us first consider the LHC bound. 
If kinematically allowed $(m_{2}< m_{{\tilde h}}/2)$, 
   the SM Higgs boson can decay to a pair of pGDMs with a branching ratio, 
\bea
BR( {\tilde h} \to \chi_2 \chi_2) = \frac{\Gamma ({\tilde h}\to \chi_2 \chi_2) }{\Gamma_{\tilde h}}. 
\eea
The CMS result on the invisible Higgs boson decay at the LHC 
   provides us with an upper bound, $BR( {\tilde h} \to \chi_2 \chi_2) \leq 0.16$ \cite{Sirunyan:2018owy}.  
In Fig.~\ref{fig:2} (left panel),  
   we show $BR( {\tilde h} \to \chi_2 \chi_2)$ as a function of the DM mass 
   (solid line) along which $\Omega_{DM} h^2 = 0.120$ is satisfied, 
    together with the CMS constraint (gray shaded).

Next, let us consider the indirect DM detection constraints.
A pair of pGDMs can annihilate into SM particles 
   whose subsequent decays produce gamma-rays. 
Such gamma-rays originating from DM pair annihilations have been searched for 
   by Fermi-LAT and MAGIC experiments. 
For a pGDM mass $\lesssim 80$ GeV, 
   a pair of pGDMs dominantly annihilates into a pair of bottom quarks. 
We interpret the upper bounds on the annihilation cross section 
   from the Fermi-LAT and MAGIC experiments into our model parameter space. 
Using the earlier result for $\lambda_{HB}$ as a function of $m_2$, 
   we calculate the pGDM pair annihilation cross section into a pair of bottom quarks. 
In Fig.~\ref{fig:2} (right panel), we show our result (solid curve), 
   along with the upper bound from the Fermi-LAT result (dashed line) and 
   the combined result by Fermi-LAT and MAGIC (dotted line). 
 The regions of $m_2 \lesssim 40$ GeV and $m_2 \simeq m_{\tilde h}/2$ are excluded.

%%%%%%%%%%%%%%%%%
\section{Conclusions}
\label{sec:conc}
%%%%%%%%%%%%%%%%%

The Higgs-portal scalar DM scenario is one of the simplest extensions of the SM with a DM candidate. 
However, this scenario is very severely constrained by 
  the null results from the direct DM detection experiments 
with nearly all of the parameter region excluded. 
The recently proposed pGDM scenario realizes the Higgs-portal scalar DM particle as a pseudo-Goldstone boson. 
Due to its Goldstone boson nature, 
  the scattering cross section  of the pGDM with a nucleon vanishes
  in the zero-momentum transfer limit, and so it evades the direct DM detection constraints.

We have proposed a pGDM scenario in the context of a gauged $B-L$ extension of the SM. 
Our model is a minimal extension of the well-known $B-L$ model with an additional $B-L$ Higgs field $\Phi_B$, and following the $B-L$ symmetry breaking, 
the Higgs sector of the model effectively realizes the pGDM scenario. 
Since the $B-L$ symmetry forbids the unwanted terms in the original pGDM model 
  which explicitly break the global U(1) symmetry and thereby spoil the Goldstone boson nature of the DM particle, 
  our model can be considered as a (gauged) ultraviolet completion of the pGDM scenario. 
Unlike the original model, the pGDM particle decays through the $B-L$ gauge interaction, 
   and the $B-L$ symmetry breaking scale is estimated to be quite high ($\sim {\cal O} (10^{11})$ GeV) in order 
   to make the pGDM lifetime sufficiently long. 
Although the model is free from the direct DM detection constraints, 
   the DM model parameter space can be constrained by 
   the LHC and gamma ray observations by Fermi-LAT and MAGIC.

Finally, in addition to the pGDM physics, 
    our model retains the salient features of the minimal $B-L$ model 
    such that the seesaw mechanism is automatically incorporated 
    and the baryon asymmetry of the universe can be reproduced through leptgenesis. 
In short, our model overcomes three major problems of the SM, 
    namely the origin of tiny neutrino masses, the nature of the DM particle, and the origin of 
    matter-antimatter asymmetry.\\

{\bf Note added:} 
While finalizing this manuscript we learned that the model we have proposed in this paper 
has very recently also been discussed by the authors of Ref.~\cite{Abe:2020iph}. 

%%%%%%%%%%%%%%%%%%%%%%%%%%%%%%%%%%%%%%%%%
\section*{Acknowledgements}
%%%%%%%%%%%%%%%%%%%%%%%%%%%%%%%%%%%%%%%%%
This work of is supported in part by the United States Department of Energy Grants DE-SC0012447 (N.O) 
  and DE-SC0013880 (D.R and Q.S) and Bartol Research Grant BART-462114 (D.R). 

%%%%%%%%%%%%%%

\end{document}